\documentclass[a4paper]{jpconf}
\usepackage{graphicx}

\newcommand{\arcsec}{$^{\prime\prime}$}

\begin{document}

\title{The Nuclear Infrared Emission of Low-Luminosity AGN}

\author{R E Mason$^1$, E Lopez-Rodriguez$^2$, C Packham$^2$, A Alonso-Herrero$^{3,}$\footnote[11]{Augusto Gonz\'alez Linares Senior Research Fellow}, N A Levenson$^4$, J Radomski$^4$, C Ramos Almeida$^{5,}$\footnote[12]{Current affiliations:
Instituto de Astrof\'\i sica de Canarias (IAC), C/V\'\i a L\'{a}ctea, s/n, E-38205, La Laguna, Tenerife, Spain; Departamento de Astrof\' isica, Universidad de La Laguna, E-38205, La Laguna, Tenerife, Spain.}, L Colina$^{6}$, M Elitzur$^{7}$, I Arextaga$^{8}$, P F Roche$^{9}$, N Oi$^{10}$}

\address{$^1$ Gemini Observatory, Northern Operations Center, 670 N. A'ohoku Place, Hilo, HI 96720, USA}
\address{$^2$ University of Florida, Department of Astronomy, 211 Bryant Space Science Center, P.O. Box 112055, Gainesville, FL 32611, USA}
\address{$^3$ Instituto de F\'{\i}sica de Cantabria, CSIC-UC, Avenida de los Castros
s/n, 39005 Santander, Spain}
\address{$^4$ Gemini Observatory, Southern Operations Center, c/o AURA, Casilla 603, La Serena, Chile}
\address{$^5$ University of Sheffield, Department of Physics \& Astronomy, S3 7RH, UK}
\address{$^6$ Departamento de Astrof\'{\i}sica,
Centro de Astrobiolog\'{\i}a (CSIC/INTA),
Instituto Nacional de T\'{e}cnica Aeroespacial,
Crta de Torrej\'{o}n a Ajalvir, km 4,
28850 Torrej\'{o}n de Ardoz, Madrid
Spain}
\address{$^{7}$ Department of Physics and Astronomy, University of Kentucky, Lexington, KY 40506, USA}
\address{$^{8}$ Instituto Nacional de Astrof\'{i}sica, \'{O}ptica y Electr\'{o}nica (INAOE), Aptdo. Postal 51 y 216, 72000 Puebla, Mexico}
\address{$^{9}$ Astrophysics, Department of Physics, University of Oxford, DWB, Keble Road, Oxford OX1 3RH, UK}
\address{$^{10}$ Department of Astronomy, School of Science, Graduate University for Advanced Studies (SOKENDAI), Mitaka, Tokyo 181-8588, Japan}

\ead{rmason@gemini.edu}

\begin{abstract}

We have obtained high-resolution mid-infrared (MIR) imaging, nuclear spectral energy distributions (SEDs) and archival Spitzer spectra for 22 low-luminosity active galactic nuclei (LLAGN; L$_{\rm bol} < 5 \times 10^{42} \; \rm erg \; s^{-1}$).  Infrared (IR) observations may advance our understanding of the accretion flows in LLAGN, the fate of the obscuring torus at low accretion rates, and, perhaps, the star formation histories of these objects. However, while comprehensively studied in higher-luminosity Seyferts and quasars, the nuclear IR properties of LLAGN have not yet been well-determined. In these proceedings we summarise the results for the LLAGN at the relatively high-luminosity, high-Eddington ratio end of the sample. Strong, compact nuclear sources are visible in the MIR images of these objects, with luminosities consistent with or slightly in execss of that predicted by the standard MIR/X-ray relation. Their broadband nuclear SEDs are diverse; some resemble typical Seyfert nuclei, while others possess less of a well-defined MIR ``dust bump''. Strong silicate emission is present in many of these objects. We speculate that this, together with high ratios of silicate strength to hydrogen column density, could suggest optically thin dust and low dust-to-gas ratios, in accordance with model predictions that LLAGN do not host a Seyfert-like obscuring torus. 

\vspace{4mm}

\end{abstract}

\section{Introduction}

\vspace{4mm}

We have been investigating the nuclear infrared (IR) properties of LLAGN, loosely defined as a low-ionisation nuclear emission region (LINER) or Seyfert galaxy with $L_{\rm bol}$ below about $10^{42}\; \rm erg\; s^{-1}$. The IR continuum emission from an LLAGN may contain contributions from a variety of processes: thermal emission from a dusty torus surrounding the central supermassive black hole (SMBH), synchrotron emission from a jet, and thermal emission from a truncated accretion disk. Dust in a narrow-line region (NLR) or associated with the surrounding stellar population may also contribute. Silicate emission or absorption features at 10 and 18 $\mu$m contain information about dust geometry and heating, and the suite of IR polycyclic aromatic hydrocarbon (PAH) bands may trace star formation in the nuclear environment. The IR wavelength regime therefore contains much information that could potentially advance our understanding of the nature and lifecycle of LLAGN. However, we do not yet have a good overview of the {\em nuclear} IR properties of these objects.

Studies of the spectral properties, luminosities, fuel supply etc. of LLAGN suggest that the standard optically thick and geometrically thin accretion disk is truncated, and replaced by a geometrically thick, radiatively inefficient accretion flow (RIAF)  interior to the disk truncation radius \cite{Yuan07,Ho09,Trump11}.  The truncation of the thin disk should shift its thermal emission peak towards longer wavelengths and indeed, \cite{Nemmen11} present models in which the bulk of the nuclear luminosity at $\sim$1--10 $\mu$m can come from the truncated disk. \cite{Nemmen11} note, though, that the available, large-aperture IR data cannot constrain the presence or properties of a truncated disk in the objects modeled. IR photometry, at higher angular resolution than is generally available in the literature,  may therefore aid our understanding of the accretion physics of LLAGN. 

In higher-luminosity AGN, the difference between types 1 and 2 is explained at least to first order by the presence of a toroidal cloud of dust and gas obscuring the AGN from certain viewing directions while permitting a direct view from others. Some models of the torus explain its existence through inflows of gas from larger scales \cite{Wada09,Schartmann10}. During periods when little material is reaching the centre of the galaxy, the torus may become thin and transparent \cite{Vollmer08}. Conversely, the torus may be part of a dusty, outflowing wind \cite{Konigl94}. The disk wind model of \cite{Elitzur06} predicts that below L$_{\rm bol} \sim 10^{42}\; \rm erg \; s^{-1} $, accretion onto the black hole can no longer sustain the outflow necessary to obscure the nucleus. In either case, low-luminosity AGN may show little nuclear obscuration and dust emission.

Observationally, there are indications that LLAGN do tend to have unobscured nuclei. For instance, nuclear UV point sources have been detected in LINERs of both types 1 and 2 \cite{Maoz05}. A decline in absorbing column density has also been observed in X-ray studies of LLAGN \cite{Zhang09}.  On the other hand, the detection of broad H$\alpha$ in polarised light in some LINERs \cite{Barth99} suggests that these objects host dust-obscured AGN. X-ray studies of single objects show that some LLAGN do have substantial absorbing columns, and the fraction of such objects may be significant \cite{Gonzalez-Martin09b}. \cite{Sturm06b} present average spectra of type 1 and 2 LINERs which suggest an extra hot dust component in the type 1s relative to the type 2s, analogous to ``conventional'' Seyferts.  The role of dust in LLAGN, and in particular whether there exists a luminosity, accretion rate, or other property at which the torus ceases to exist, remains unclear. A search for the IR signatures of the torus -- thermal emission and silicate emission and absorption features -- promises a better understanding of these issues. 
 
Until recently,  high-resolution imaging at $\lambda \sim 10 \; \mu$m existed only for a handful of relatively bright, well-known objects. More recently, \cite{Asmus11} have presented ground-based MIR imaging of a number of LLAGN, with several new detections (see also contributions from Asmus and Fernandez-Ontiveros in this volume). NIR imaging is also available in the literature, but it has rarely been considered in the context of the multi-wavelength emission of LLAGN. Published LLAGN SED compilations contain little or no high-resolution IR data \cite{Ho99,Eracleous10a}.

\begin{figure}[th]
\includegraphics[scale=0.86,  angle=270, clip, trim=75 95 270 170]{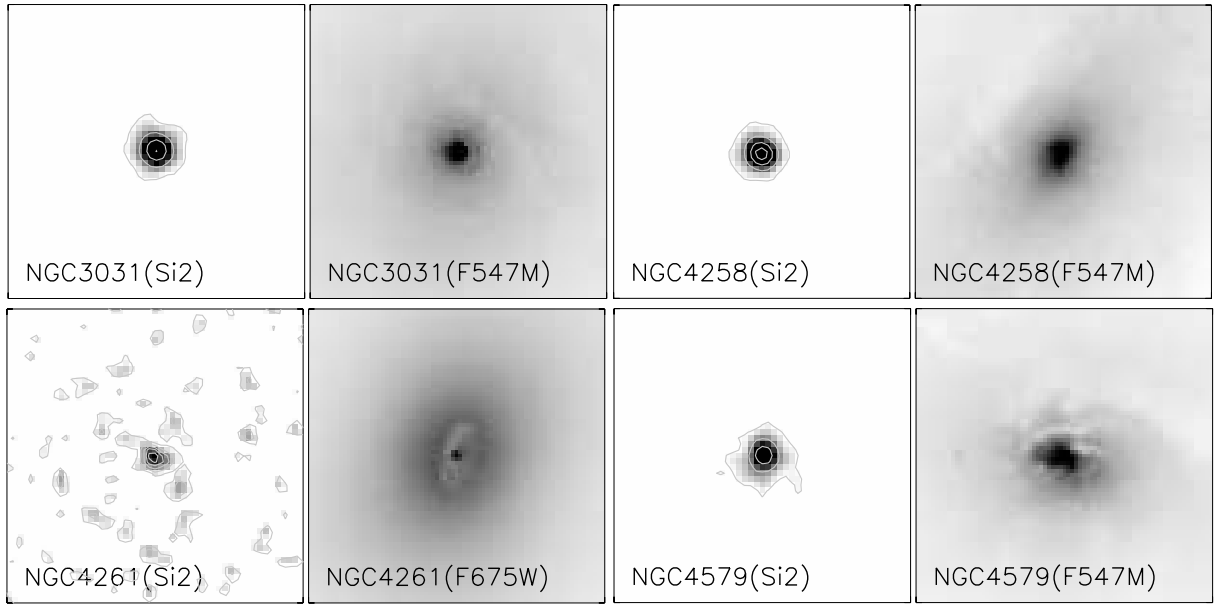}
\caption{\label{fig:images} Images of four of the LLAGN at 8.8~$\mu$m, along with HST optical images. N is up and E left on the images, and 5\arcsec $\times$ 5\arcsec\ regions are shown. }
\end{figure}

To establish the overall nuclear IR properties of LLAGN, we have acquired high-resolution, Gemini MIR  imaging of 20  IR-faint nuclei with L$_{\rm bol} < 5 \times 10^{42} \; \rm erg \; s^{-1}$. We combine these data with published measurements in the IR and at other wavelengths to produce nuclear spectral energy distributions (SEDs) for these objects. In terms of high spatial resolution IR data, these SEDs are by far the most detailed yet available. Finally, we also analyse archival Spitzer low-resolution spectroscopy for the 18/22 galaxies with available data. In these proceedings we briefly describe the morphological, spectral and broadband characteristics of the 12 galaxies with relatively high Eddington ratios  (log $\rm L_{bol} / L_{Edd} > -4.6$; cf median log $\rm L_{bol} / L_{Edd} \sim -6$ for the LINERs and log $\rm L_{bol} / L_{Edd} \sim -4$ for the Seyferts in the Palomar galaxy sample, \cite{Ho09}). Results for the full sample are presented in a forthcoming paper (Mason et al., submitted).

\section{Morphology and Silicate Features}

\vspace{4mm}

Example 8.8~$\mu$m images of four of the LLAGN are presented in Figure \ref{fig:images}. Unlike the LLAGN of lower Eddington ratio in the sample, the images of these relatively high-$\rm L_{bol} / L_{Edd}$ objects tend to be dominated by compact nuclei. In higher-luminosity AGN, such as QSOs and "conventional" Seyferts, the torus manifests itself as a bright, pointlike MIR source. To investigate whether the compact MIR nuclei in these LLAGN are evidence of torus emission, we use the well-known MIR/X-ray relation \cite{Gandhi09,Levenson09} to predict the amount of MIR emission expected from a dusty torus reprocessing high-energy emission from an active nucleus. We fit the MIR and X-ray luminosities of the combined AGN samples of \cite{Gandhi09,Levenson09} and extrapolate the fit to the luminosities of the LLAGN. 

\begin{figure}[th]
\includegraphics[scale=0.81,  angle=270, clip, trim=100 115 290 110]{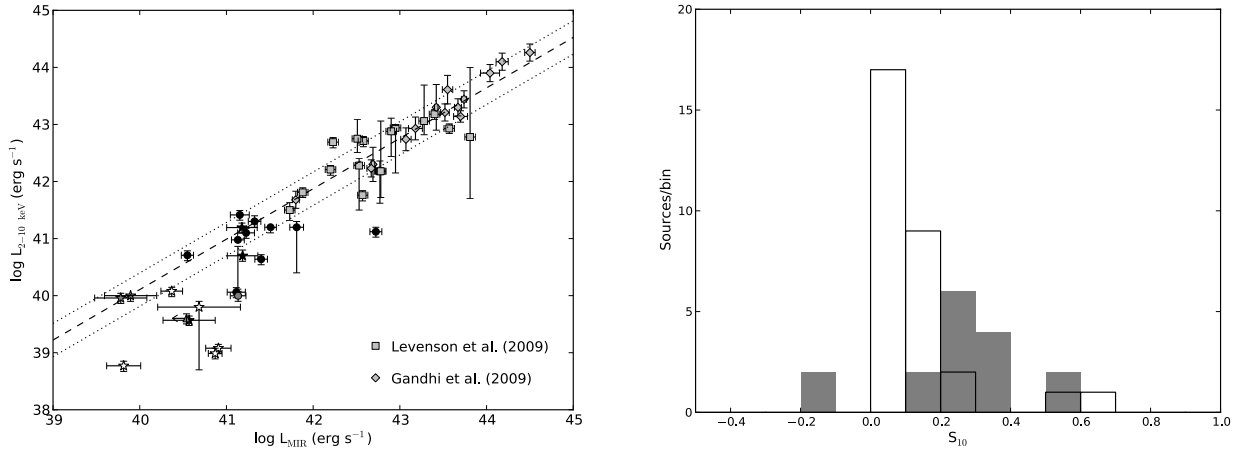}
\caption{\label{fig:mirx} Left: X-ray vs MIR luminosities for the LLAGN, as well as the Seyfert and QSO samples of \cite{Gandhi09,Levenson09}. Solid black symbols represent the LLAGN discussed in this paper. Right: Strength of the 10 $\mu$m silicate feature in the LLAGN with log $\rm L_{bol} / L_{Edd} > -4.6$ (grey) and in the type 1-1.5 Seyferts of \cite{Thompson09}. Positive values of S$_{10}$ indicate emission. For clarity, NGC~7479 from the LLAGN sample and UGC~5101 from the Seyfert 1 sample, which have deep absorption features ($S_{10}$ = -2.19 and -1.52 respectively), are not shown.}
\end{figure}

A MIR deficit in the MIR/X-ray plot would be consistent with suggestions that a Seyfert-like torus does not exist in the LLAGN. However, the high-$\rm L_{bol} / L_{Edd}$ LLAGN lie close to the MIR/X-ray relation or exhibit excess MIR emission. In most cases the MIR excess is only significant at the 1-2$\sigma$ level, but it is noteworthy that only two  of the 12 high-$\rm L_{bol} / L_{Edd}$ objects lie to the MIR deficit side of the fit. Similar results are reported by \cite{Asmus11}; while finding that the MIR/X-ray relation is formally unchanged down to L$\sim10^{41} \rm erg \; s^{-1}$, their LLAGN are offset from the relation by about 0.3 dex to the MIR excess side.

 \cite{Horst08} find that, even in a sample of Seyfert galaxies with pointlike nuclei in high-resolution MIR imaging, objects with FWHM $ \rm >560  \rm \; r_{sub}$ (the dust sublimation radius) are systematically offset from the MIR/X-ray relation for ``well-resolved'' AGN. They attribute this to contamination from nuclear MIR sources other than the torus. All of the LLAGN in this sample have FWHM $ \rm >560 \; r_{sub}$, so a contribution from unresolved, non-torus emission sources -- nuclear star formation, synchrotron radiation, truncated accretion disk, etc. -- is possible. We return to this issue in \S\ref{discuss}. 

9/12 of these LLAGN have been observed with Spitzer's InfraRed Spectrograph. Strikingly, all but two of them exhibit silicate emission bands, in many cases very strong. Figure \ref{fig:mirx} compares the strength of the 10~$\mu$m silicate feature in the LLAGN with the Seyfert 1-1.5 sample of \cite{Thompson09}. The LLAGN sample is neither complete nor unbiased, and Figure \ref{fig:mirx} is not intended to indicate that the distributions of S$_{10}$ differ systematically between the various AGN types. However, it does demonstrate that the 10~$\mu$m silicate features found in these particular AGN tend to be comparable to the stronger emission features in the Thompson et al. Seyfert 1 sample.  As our sample selection was partly based on the existence of previous, large-aperture photometry, objects with strong silicate emission features could in principle have been more likely to be included in this study. As the flux increases by less than a factor of two across the 10~$\mu$m feature, though, it is not clear that this would be a significant effect. In \S\ref{discuss} we suggest that the strong silicate emission features may represent a particular stage in the evolution of the torus.

\section{Spectral Energy Distributions}
\label{sed}

\vspace{4mm}

\begin{figure}[t]
\includegraphics[scale=0.9,  angle=270, clip, trim=85 95 145 210]{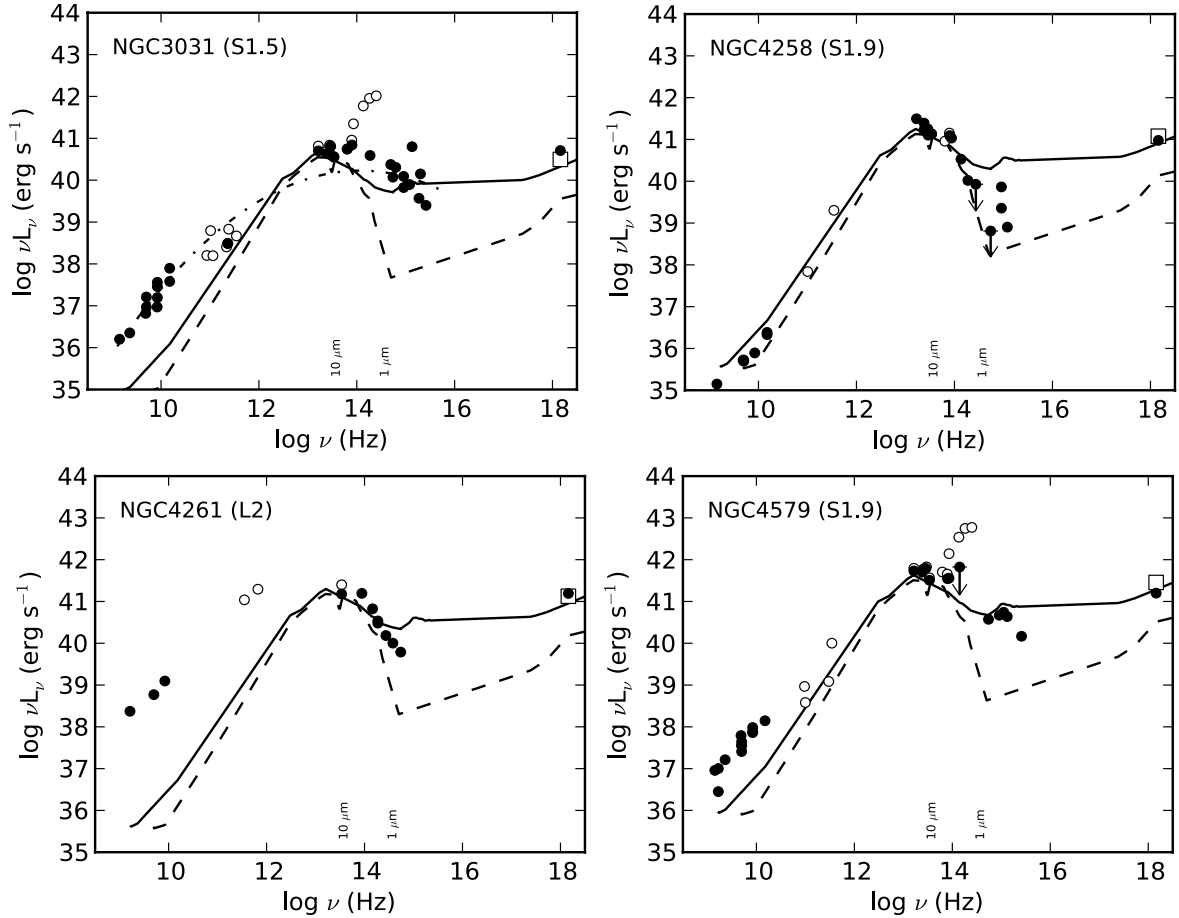}
\caption{\label{fig:seds} Example SEDs of the LLAGN. Solid points indicate high-resolution data representing the AGN emission. The solid and dashed lines denote the mean type 1 and 2 Seyfert SEDs of \cite{Prieto10}, respectively. }
\end{figure}

Example high-resolution radio -- X-ray SEDs of the  high-$\rm L_{bol} / L_{Edd}$  LLAGN are shown in Figure \ref{fig:seds}. The SEDs are quite diverse. If they have one distinguishing feature, it is that many of them are more radio-loud than the average Seyfert galaxy. This is expected from previous work showing that radio loudness increases with decreasing Eddington ratio (e.g. \cite{Sikora07}).

Some of the SEDs -- such as those of NGC~4258,  NGC~4261 and NGC~4579  -- show a well-defined, Seyfert-like peak at MIR wavelengths. The MIR - NIR/optical spectral slopes in these objects are within the range bracketed by the  Seyfert 1 and 2 templates of \cite{Prieto10}. Amongst these objects, nuclei of the same AGN type do not necessarily have the same spectral slope; for instance, the type 1.9 Seyferts NGC~4258 and NGC~4579. However, if intermediate-type LLAGN have similar nuclei to more luminous intermediate-type objects, then a variety of spectral shapes is to be expected; \cite{Alonso-Herrero03} find a wide range of spectral indices in Seyferts of types 1.8 and 1.9 in the CfA sample. Judged solely by their SEDs, there is little indication that these objects are anything but ``scaled-down'' Seyferts,  complete with a MIR ``dust bump'', albeit with extra radio emission. 

A few of the galaxies -- NGC~3031, for example -- have rather flat MIR -- NIR/optical SEDs compared to the Seyfert templates, and a less well-defined MIR peak. These are also characteristics of the radio-loud, low-Eddington ratio objects in the sample, whose IR emission we find to be dominated by the jets (Mason et al., submitted). 
In some objects, such as NGC~1097, the ``excess'' NIR/optical emission is probably related to a young, massive nuclear star cluster \cite{StorchiBergmann05,Mason07}. 

Overall, many of the high-Eddington ratio LLAGN have SEDs that broadly resemble those of conventional Seyferts, although often with enhanced radio emission. In some cases, though, the MIR -- NIR/optical SED slopes are flatter than expected for Seyferts, which hints at possible stellar and/or jet contributions in the IR. Other sources, such as a truncated accretion disk, may also influence the IR properties of the SEDs. Detailed modeling of the SEDs with jet, RIAF, thin disk and torus components is beyond the scope of the present work, but in \S\ref{discuss} we suggest a scenario which is consistent with the morphological, spectral and broadband SED characteristics of these LLAGN.

\section{Discussion}
\label{discuss}

\vspace{4mm}

The new, high-resolution IR imaging and SEDs presented here represent the first step towards establishing the nuclear IR properties of a significant number of LLAGN. At the median distance of the galaxies studied, 16.8 Mpc, the AGN is well isolated in objects with log L$_{2-10 \rm \; keV} > 40.5 \rm \; erg \; s^{-1}$ ($\rm log \; L_{bol} > 41.8 \rm \; erg \; s^{-1}$). Deeper observations would allow a cleaner detection of the pointlike central engine in less luminous or more distant objects, but the cost in observing time would be large with current facilities.

The log $\rm L_{bol} / L_{Edd} > -4.6$ nuclei discussed in these proceedings, which lie at the high luminosity end of the sample, tend to have prominent nuclear point sources in the MIR. They often have strong silicate emission features compared to those typically observed in Seyfert 1 nuclei, and their MIR luminosities are consistent with or slightly in excess of those predicted by the standard MIR/X-ray relation. The nuclear broadband SEDs of these objects are rather mixed. Some are essentially indistinguishable from those of ``conventional'' Seyferts, many have excess radio emission compared to higher-luminosity Seyferts, and a few have unusually flat MIR - NIR/optical slopes. 

\begin{figure}[t]
\includegraphics[scale=0.8,  clip, trim=35 200 50 200]{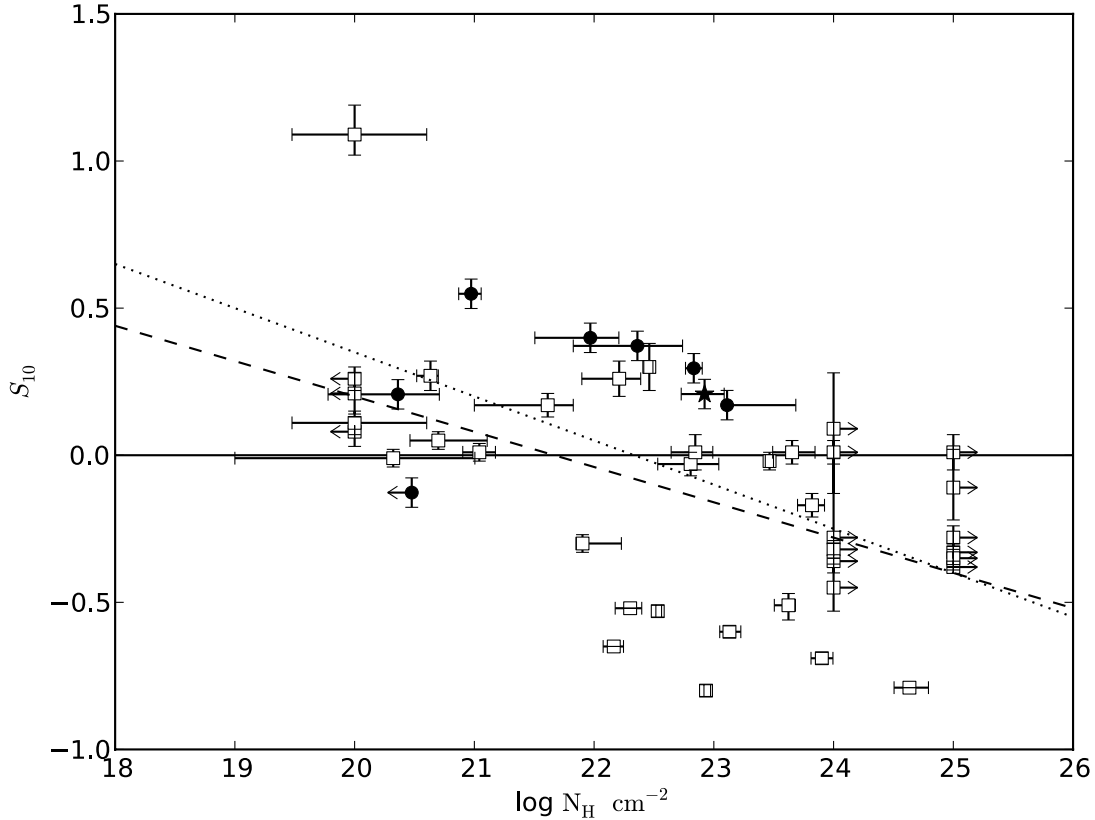}
\caption{\label{NH} Strength of the 10 $\mu$m silicate feature vs X-ray column density for the high-Eddington ratio LLAGN (black). Square symbols indicate the Seyfert galaxies of \cite{Shi06}. The dashed line shows Shi et al.'s fits to their Seyferts, the dotted line their fit to their whole sample (mostly comprised of Seyferts and various classes of quasar). Positive values of $S_{10}$ indicate emission, negative values absorption. As in Figure \ref{fig:mirx}, NGC~7479, which has a deep absorption feature ($S_{10} = -2.19$), is omitted for clarity.}
\end{figure}

Considered in isolation, these characteristics offer no compelling evidence that the torus is absent in these LLAGN or that it differs from that observed in higher-luminosity Seyferts. However, we do find some indications that the IR emission may not arise in the standard, Seyfert-like torus of the AGN unified model. In Figure \ref{NH}, we plot the strength of the 10~$\mu$m silicate feature against HI column density, for both the Seyfert galaxies of \cite{Shi06} and the high-$\rm L_{bol} / L_{Edd}$ LLAGN. In the Seyferts of \cite{Shi06}, S$_{10}$ and N$_{\rm H}$ are loosely correlated. The scatter presumably reflects the fact that the observed $S_{10}$ is likely a complicated function of the precise arrangement of clouds in the torus \cite{Honig10b}, and may also indicate a contribution from absorption in the host galaxy \cite{Roche07,Deo09,Alonso-Herrero11}. Many of the high-Eddington ratio LLAGN lie on the upper envelope of the Seyfert points, with relatively high values of $S_{10}$ per unit N$_{H}$.  One possible explanation is that the dust-to-gas ratio is lower in these particular objects than in most Seyfert galaxies. This may be expected if the torus is an optically thick region of an accretion disk wind that becomes less powerful at low accretion rates \cite{Elitzur09}, when the amount of material reaching the dust sublimation radius and able to form dust grains is reduced. In this case we may expect either a torus with fewer clouds and a higher probability of observing a hot, directly-illuminated cloud face, or simply optically thin dust emission. Both of these configurations would cause relatively strong silicate emission features, consistent with our finding that S$_{10}$ in many of these high-Eddington ratio LLAGN is strong compared to that in typical type 1 Seyferts (Figure \ref{fig:mirx}).

A contribution to the observed silicate features from circumstellar shells in the host galaxy is also possible \cite{Bressan06,Buson09}. As the luminosity of the central engine diminishes, the relative strength of features arising in the surrounding stellar population will increase. This could also account for changes in $S_{10}$/N$_{\rm H}$ among LLAGN.

The high-Eddington ratio LLAGN have at least as much MIR emission as predicted by the MIR/X-ray relation for Seyferts and quasars. If the torus in these LLAGN does indeed contain less dust, the IR continuum emission must be produced by some other mechanism. A contribution from synchrotron radiation is likely, and may explain the unusually flat MIR -NIR/optical slopes observed in some of the SEDs. Another possibility is that some of the IR emission comes from a truncated accretion disk. In the models of \cite{Nemmen11}, the disk emission peaks between 1 -- 10~$\mu$m and can account for essentially all the luminosity at these wavelengths. Emission associated with nuclear star clusters, such as that known to exist in NGC~1097 \cite{StorchiBergmann05} may also play a r\^{o}le. However, \cite{Asmus11} find that star formation contributes $<$30\% of the small-scale 12~$\mu$m flux in most LLAGN of comparable luminosity to those studied in this paper. Detailed modelling of the LLAGN, taking advantage of the new constraints from the well-sampled IR SEDs presented here, will provide more insight into the processes responsible for the IR emission of LLAGN and the role of the torus in these objects.

\ack

This paper is based on observations obtained at the Gemini Observatory, which is operated by the
Association of Universities for Research in Astronomy, Inc., under a cooperative agreement
with the NSF on behalf of the Gemini partnership: the National Science Foundation (United
States), the Science and Technology Facilities Council (United Kingdom), the
National Research Council (Canada), CONICYT (Chile), the Australian Research Council
(Australia), Minist\'{e}rio da Ci\^{e}ncia e Tecnologia (Brazil) and SECYT (Argentina). A.A.-H. and L. C. acknowledge support from the Spanish
Plan Nacional de Astronom\'{\i}a y Astrof\'{\i}sica under grants
AYA2009-05705-E and AYA2010-21161-C02-1.

\vspace*{4mm}


\end{document}